\documentclass{llncs}
\usepackage{makeidx}  
\usepackage{graphicx}
\begin{document}

\title{Double Calorimetry System in JUNO}
\author{Miao He \\ on behalf of the JUNO collaboration}
\institute{Institute of High Energy Physics, Beijing\\
\email{hem@ihep.ac.cn}}


\maketitle              

\begin{abstract}
The Jiangmen Underground Neutrino Observatory (JUNO) is a multipurpose neutrino-oscillation experiment, with a 20 kiloton liquid scintillator detector of unprecedented 3\% energy resolution (at 1 MeV) at 700-meter deep underground. There are ~18,000 20-inch photomultiplier tubes (PMTs) in the central detector with an optical coverage greater than 75\%. Control of the systematics of the energy response is crucial to archive the designed energy resolution as well as to reach 1\% precision of the absolute energy scale. The detected number of photoelectrons in each PMT differs by two orders of magnitude in the reactor antineutrino energy range in such a large detector, which is a challenge to the single channel charge measurement. JUNO has approved a new small-PMT system, including 25,000 3-inch PMTs, installed alternately with 20-inch PMTs. The individual 3-inch PMT receives mostly single photoelectrons, which provides a unique way to calibrate the energy response of the 20-inch PMT system by a photon-counting technology. Besides, the small-PMT system naturally extends the dynamic range of the energy measurement to help the high-energy physics, such as cosmic muons and atmospheric neutrinos. We will present the physics concept of this double calorimetry, the design and implementation of the 3-inch PMT and its readout electronics system.
\end{abstract}
\section{Introduction}

JUNO~\cite{juno.cdr} is a neutrino experiment under construction in southern China, $\sim$53~km from two powerful nuclear plants, Yangjiang and Taishan. The detector target contains 20~kiloton liquid scintillator with 3\% energy resolution (at 1 MeV). It is located 700~m underground with a cosmic muon flux of 0.0030~Hz~m$^{-2}$. The primary goal of JUNO is to determine the neutrino mass hierarchy using reactor antineutrinos by the interference between two oscillation frequency components driven by $\Delta m^{2}_{31}$ and $\Delta m^{2}_{32}$, respectively. On the other hand, the precise measurement of antineutrino spectra allows JUNO to be the first experiment to measure solar and atmospheric mass splitting simultaneously, with better than 1\% precision of $\theta_{12}$, $\Delta m^{2}_{21}$ and $\Delta m^{2}_{31}$ (or $\Delta m^{2}_{32}$). In addition, the large detector volume, good energy resolution and very low radioactive background allow more physics possibilities, including supernova neutrinos, terrestrial neutrinos, solar neutrinos and exotic searches~\cite{juno.yb}.

\begin{figure}
\centering
\includegraphics[width=0.9\columnwidth]{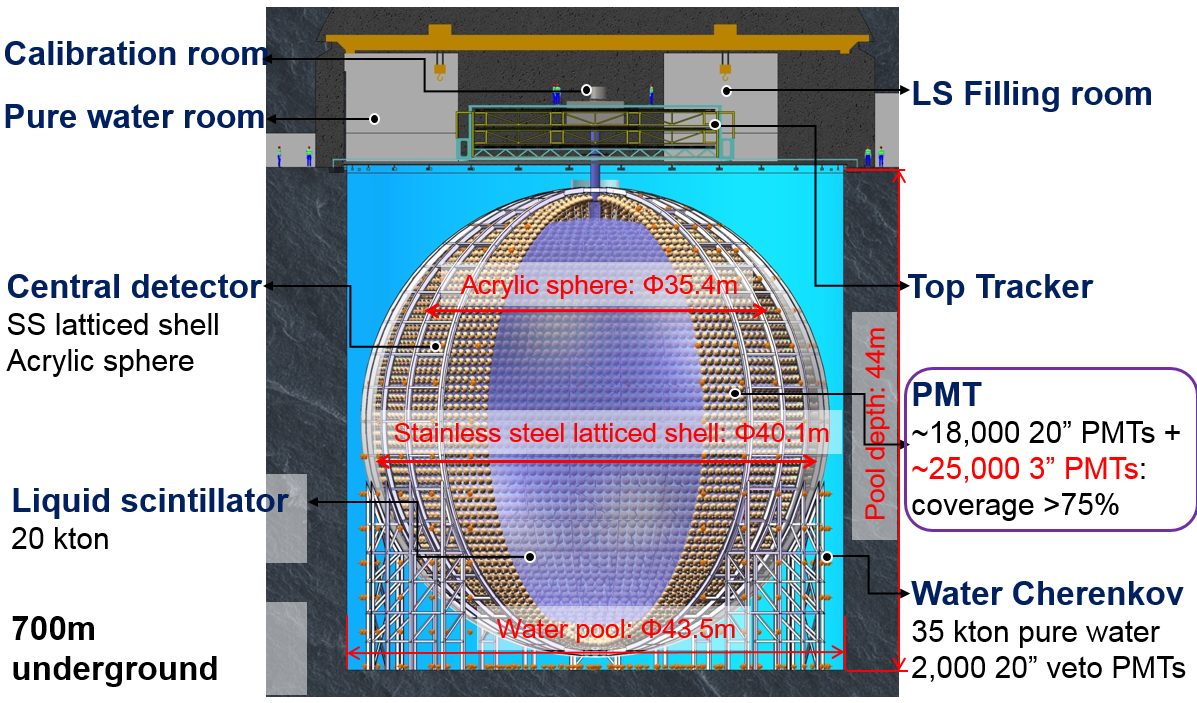}
\caption{JUNO detector design\label{fig.det}}
\end{figure}

The schematic design is shown in Fig.~\ref{fig.det}. There is an acrylic sphere with an inner diameter of 35.4~m, which is the container of the liquid scintillator, and supported by a stainless steel latticed shell at a diameter of 40.1~m with 590 connecting bars. There are $\sim$18,000 20-inch photomultiplier tubes (PMTs) and $\sim$25,000 3-inch PMTs installed on the stainless steel shell, watching inward on the light generated by the interaction of neutrinos. All of the detector components are immersed in a large pool, filled with 35~kiloton pure water. The water pool also serves as a Cherenkov detector after equipping with 2,000 20-inch PMTs to tag cosmic muons. A redundant muon veto system made of plastic scintillators is deployed on top of the pool.

\section{Double calorimetry}

Determination of the mass hierarchy requires precision measurement of the energy spectrum to separate $\Delta m^{2}_{31}$ and $\Delta m^{2}_{32}$, and the sensitivity of the mass hierarchy heavily depends on the detector energy resolution~\cite{liyf.mh}. JUNO aims for 3\%/$\sqrt{E(\rm{MeV})}$ resolution, which needs very large number of detected photons and strict control of the systematics. High light yield, high transparency liquid scintillator and high quantum efficiency 20-inch PMTs with 75\% optical coverage give $>$1,200 photoelectrons at 1~MeV with a statistical fluctuation of 2.9\%, which leaves a room of $<$1\% for the systematic uncertainty.

Unfortunately, the 20-inch PMT is too large to see large variation of the detected number of photoelectrons by two orders of magnitude in the reactor antineutrino energy range in such a large detector. It is a challenge to calibrate the non-linear response of the single channel charge measurement to sub-percent level~\cite{dyb.nl}. Moreover, when the event vertex differs in the detector volume, the charge range at a single PMT also changes. Therefore, the nonlinearity of the single channel contributes to a non-uniformity of the detector and thus deteriorates the energy resolution.

A small-PMT system was proposed in 2014 to install up to 36,000 3-inch PMTs in the gap between 20-inch PMTs. The photocathode area is $\sim$1/50 compared to the large PMT, and more than 98\% of small PMTs only detect single photoelectron in the reactor antineutrino energy range in the JUNO detector according a Monte Calor simulation. Therefore, the total charge of the small-PMT system can be obtained with a photon-counting technology and thus there is almost no charge nonlinearity. This feature provides a unique way to calibrate the energy response of the 20-inch PMT system. Combination of the large and the small PMTs system constitutes a {\it double calorimetry} to control both stochastic and non-stochastic effects. Besides, the small-PMT system naturally extends the dynamic range of the energy measurement to help the high-energy physics, such as cosmic muons and atmospheric neutrinos. The physics concept and the system design of the double calorimetry was approved sequentially by the JUNO collaboration in 2015 and 2016.

\section{Small PMT system}

\begin{figure}
\centering
\includegraphics[width=0.9\columnwidth]{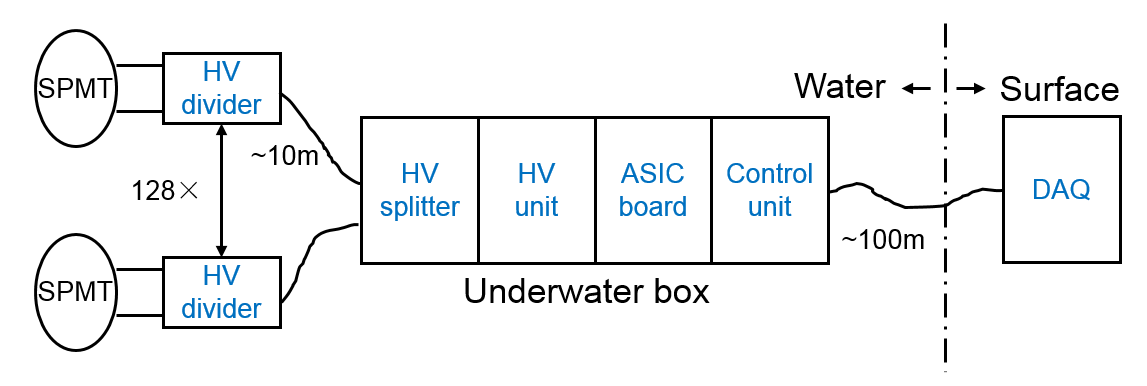}
\caption{Schematic design of the small PMT system.\label{fig.spmt}}
\end{figure}

The small-PMT system consists of PMTs and the readout electronics. As shown in Fig.~\ref{fig.spmt}, a group of 128 small PMTs (SPMTs) with HV divider and water proofing is connected to an underwater box, which is the container of the readout electronics system. A $\sim$100~m cable transmits power and data to the surface.

An international bidding of PMTs was organized in May, 2017. In the end, Hainan Zhanchuang (HZC) photonics, a Chinese industry who introduced the production line and technologies from PHOTONIS, was chosen to be the supplier. HZC will produce 25,000 3-inch PMTs (XP72B22) with 1,000 spares for JUNO in the next two years. XP72B22 is an upgrade of previous XP72B20, with dedicated R$\&$D of better timing based on the requirements of JUNO. HZC will also collaborate with JUNO and produce HV dividers and do the water proofing for all PMTs.

Performances of XP72B22 are listed in Table.~\ref{table.hzc}. All of them meet JUNO's requirements. In particular, the single photon detection is critical for JUNO since the small PMT works mostly in the photon-counting mode, and the resolutions of the single photoelectron of 5 samples from HZC were measured to be (35$\pm$2)\%, showing very good uniformity.

\begin{table}
\caption{Highlight of HZC's XP72B22 performances\label{table.hzc}}
\begin{center}
\begin{tabular}{c|c} \hline
Parameters & HZC's response \\
\hline
Photon detection efficiency@420~nm & 24\% \\
TTS(FWHM) of single photoelectron & $<$5ns \\
P/V ratio of single photoelectron & 3 \\
Single photoelectron resolution & 35\% \\
Dark rate @ 0.25~PE & 1,000~Hz \\
Quantum efficiency uniformity & $<$30\% in $\Phi$60~mm \\
Pre/after pulse charge ratio & $<$5\%/$<$15\% \\
Nonlinearity & $<$10\%@1-100~PE \\
Radioactivity & $^{238}$U$<$400~ppb, $^{232}$Th$<$400~ppb, $^{40}$K$<$200~ppb \\
\hline
\end{tabular}
\end{center}
\end{table}

The HV unit and the control unit~\cite{gcu} are expected to be the same as the 20-inch PMT system. A a multichannel front-end ASIC CATIROC was chosen to read out the charge and time information from the small PMT with almost no dead time. Details of CATIROC can be found in Ref.~\cite{catiroc}. The preliminary design of the underwater box includes a commercial stainless steel tube with two caps. A multichannel connector between PMTs and the box is under consideration for easier installation with reasonable price.

\section{Conclusion}

Energy resolution is the key of the reactor antineutrino spectrum measurement, for the determination of the neutrino mass hierarchy. A small-PMT system was proposed to work with the large-PMT system as a double calorimetry, in order to control both the stochastic and non-stochastic effects, and thus to improve the energy resolution. JUNO has signed a contract with HZC in May 2017, and is going to install 25,000 3-inch PMTs in the central detector. The mass production and testing of PMTs are in preparation, and R$\&$D of the readout electronics and the underwater box is ongoing. The system is expected to be ready in 2020, and starts taking data with the full detector.

This work is supported by the National Natural Science Foundation of China No. 11575226, and the Strategic Priority Research Program of the Chinese Academy of Sciences, Grant No. XDA10011200.

%
%

\end{document}